\documentclass[aps,preprint,superscriptaddress,amsmath,amssymb,floatfix,showpacs]{revtex4-1}
\usepackage{times}
\usepackage{graphicx,graphics,color}
\usepackage{color}

\begin{document}

\begin{center}
{\large\bf 
Site-Selective Electronic Correlation in $\alpha$-Plutonium Metal
}
\\[0.5cm]


Jian-Xin Zhu,$^{1}$ R. C. Albers,$^{1}$ K.  Haule,$^{2}$ G. Kotliar,$^{2}$ and J. M. Wills$^1$ \\

$^1$ {\em Theoretical Division, Los Alamos National Laboratory,
Los Alamos, New Mexico 87545, USA}\\

$^2$ {\em Department of Physics and Astronomy, Rutgers University, Piscataway, New Jersey 08854, USA} \\


%
%
%
%
%


\end{center}

\vspace{0.5cm}
{\bf 
\noindent \underline{Abstract}\\
An understanding of the  phase diagram of elemental plutonium (Pu) must include both the effects of the strong directional bonding and the high density of states of the Pu $5f$ electrons, as well as how that bonding weakens under the influence of strong electronic correlations.  
 We present for the first time electronic-structure calculations of the full 16-atom per unit cell $\alpha$-phase structure  within the framework of density functional theory (DFT)  together with dynamical mean-field theory (DMFT).
Our calculations demonstrate that Pu atoms sitting on different sites within the $\alpha$-Pu crystal structure have a strongly varying site dependence of the localization-delocalization correlation effects of their $5f$ electrons and a corresponding effect on the bonding and electronic properties of this complicated metal.
In short, $\alpha$-Pu has  the capacity to simultaneously have multiple degrees of electron localization/delocalization of Pu $5f$ electrons within a pure single-element material.
}

\maketitle

\newpage

Pure plutonium has long been considered to be one of the most exotic elemental solids in the periodic table with respect to its very unusual physics properties, crystal structure, and metallurgy~\cite{OJWick67,DAYoung91,NGCooper00,RCAlbers01,SSHecker04,RCAlbers07}, including a phase diagram with six allotropic phases, with the low-temperature monoclinic $\alpha$-phase stable up to 395 K, and the technologically important face-center-cubic (fcc) $\delta$-phase stable between 592 K and 724 K. Importantly, for practical purposes, small amounts of other alloying elements have been shown to stabilize the $\delta$-phase at room temperature.  In both the pure material and also in these alloys the $\delta$-phase transformation of Pu exhibits  a significant atomic volume expansion (about $25\%$) relative to the $\alpha$-phase.  This $\delta$-phase has a variety of unusual and anomalous physical and mechanical properties~\cite{NGCooper00} that are believed to be caused by a change of the 5$f$ electronic orbitals from having an itinerant (metallic) to a localized behavior. This occurs for plutonium,  since this element is at the boundary between the light actinides that have itinerant  5$f$ electrons and the heavy actinides with localized 5$f$ electrons~\cite{RCAlbers01}. The actinides are the only materials in the periodic table that exhibit this type of crossover in the middle of a row.

The $\delta$-phase of Pu has frequently been an exciting topic of discussion within the condensed matter physics community, since it has a simple fcc crystal structure and strong electronic correlation effects, which
quench the expected magnetism that is typically expected for narrow-band materials. 
The simplest theoretical treatment, band-structure calculations within the local density approximation (LDA) or the generalized gradient approximation (GGA), 
fails to predict the equilibrium volume of the nonmagnetic $\delta$-phase of Pu~\cite{PSoderlind94,MDJones00}
due to the strong electronic correlations in Pu.
Several research groups
have therefore applied the LDA+$U$ method in order to include more $f$-$f$ correlation 
energy. By adjusting the on-site Coulomb repulsion energy ($U$) 
appropriately, it has been possible to fit to the experimental 
$\delta$-phase volume.  
However, earlier LDA+$U$ calculations also indicated an instability of 
$\delta$-Pu toward an antiferromagnetic ground state~\cite{JBouchet00,DLPrice00,SYSavrasov00}.  
Later calculations based on the  ``fully-localized-limit''~\cite{AOShorikov05}
or ``around-mean-field''~\cite{ABShick05,ABShick06,FCricchio08}   LSDA+$U$ with the spin-orbit interaction method were able to obtain the experimentally observed quenching of the spin and orbital magnetism  by using a closed-shell Pu 5$f^{6}$ configuration as well as  the correct $\delta$-phase 
volume. However, due to its inherently static treatment of electronic correlation,
 the LDA+$U$ approach fails to reproduce a major 5$f$-character  quasiparticle peak near the Fermi energy as observed by photoemission spectroscopy on $\delta$-Pu~\cite{AJArko00,LHavela02}. To reproduce the spectral properties, in addition to other thermodynamical ones, improved approaches are needed to capture the quantum fluctuation effects of electronic correlation.   The dynamical mean-field theory is a many-body  
 technique that enables the treatments of  band- and atomic-like aspects of electronic states on the same footing~\cite{ThPruschke95,AGeorges96}. Merging this with LDA-based methods, LDA+DMFT method~\cite{VIAnisimov97,GKotliar06} provides a new calculation scheme for strongly correlated electronic materials. Within the LDA+DMFT approach~\cite{GKotliar06,SYSavrasov01},  
the origin of substantial volume expansion of $\delta$-Pu has been explained in terms
of a competition between Coulomb repulsion and kinetic energy.  Later LDA+DMFT calculations with various impurity solvers  have not only reaffirmed the absence of magnetism in $\delta$-Pu~\cite{LVPourovskii06,JHShim:2007}, but have also reproduced the quasiparticle peak near the Fermi energy~\cite{LVPourovskii:2007,JHShim:2007,AShick07,JXZhu:2007,CAMarianetti:2008,EGorelov:2010}.  These intensive studies have established firmly the dual character of the $5f$ electronic states in Pu. 

Although much attention has been focused on $\delta$-Pu (for reasons given above), in this Communication,  we show that $\alpha$-Pu is perhaps even more interesting.  Surprisingly, for a pure single-element material, our LDA+DMFT calculations show that different Pu atoms in 
$\alpha$-Pu have different amounts of electronic correlation effects that range from itinerant-like to much more strongly localized.  
Our calculations demonstrate that this is caused by the effects of dissimilar near-neighbor atomic coordinations for the different sites on the hybridization strengths of the Pu 5$f$ electrons to the other conduction electrons, which are subsequently reflected in the effects of correlation on the site-projected densities of states. These results show that, if the element Pu is a ``crossover'' element, one might consider the $\alpha$-phase as a ``crossover'' phase, with mixed electronic correlation among its different atoms that eventually locks into the $\delta$-phase, where every Pu atom is strongly correlated.

\begin{figure}[b!]
\centering\includegraphics[
width=1.0\linewidth,clip]{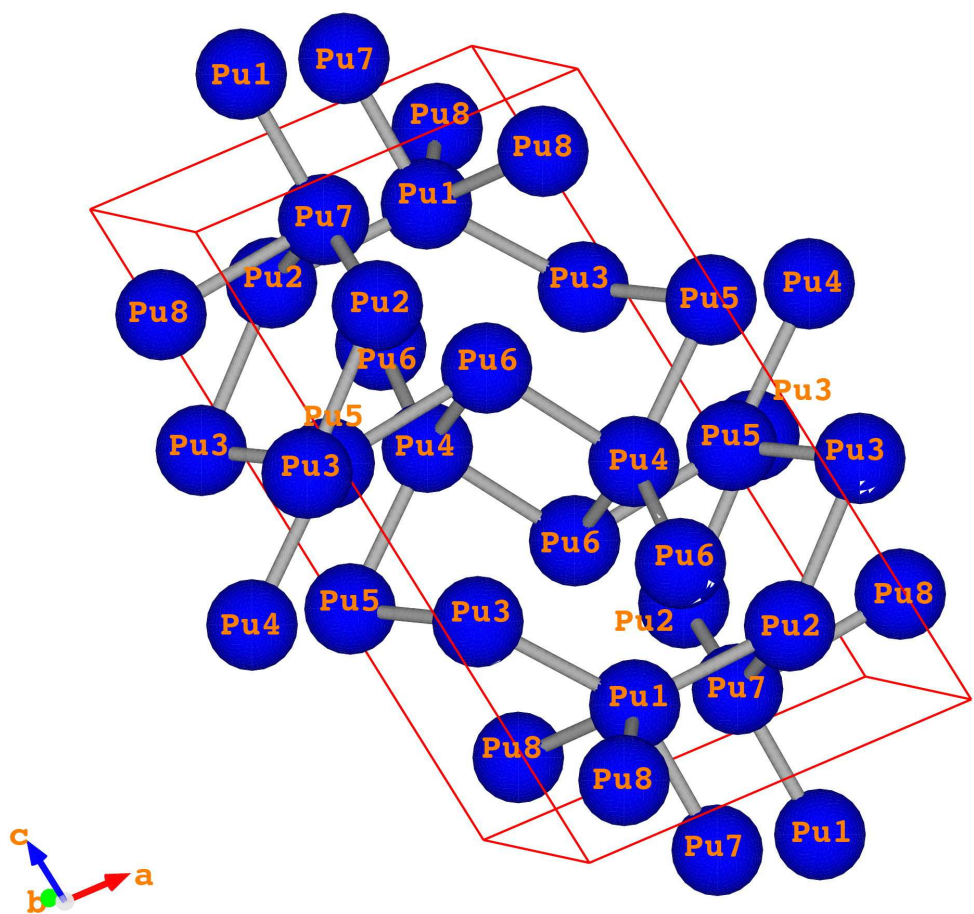}
\caption{{\bf Crystal structure of $\alpha$-Pu.} The 16 atoms are grouped into 8 nonequivalent atomic types.
}
\label{FIG:crystal}
\end{figure}

\noindent
{\bf Results}
\\
{\bf Crystal structure and bond lengths.} So far, almost all of the more sophisticated recent calculations have been done for predominantly the fcc $\delta$-phase, because this crystal structure has only one Pu atom per unit cell and is therefore amenable to better but more computationally intensive techniques (such as LDA+DMFT) to handle the electronic correlations.  
However, the equilibrium phase at room temperature and below is actually the $\alpha$-phase of Pu.  This phase has a long history of controversy.  Its crystal structure is very complex.  
As shown in Fig.~\ref{FIG:crystal}, $\alpha$-Pu has a low symmetry monoclinic structure with space group $P21/m$. It involves 16 atoms per unit cell with eight unique atomic sites~\cite{WZachariasen:1963,FJEspinosa:2001}. At 294 K, the unit cell has dimensions of $a=6.183$  \AA, $b=4.822$ \AA, $c=10.963$ \AA, and $\beta=101.79^\circ$. All the atoms are located at either $(x,1/4,z)$ or $(-x,3/4,-z)$ positions, but they have a very complicated and distorted nearest-neighbor arrangement of Pu-Pu bonds, as we shall describe below.  A simple intuitive picture can be derived from a tight-binding description, which is natural for a narrow band system like Pu.  Since the $f$ electrons dominate the bonding properties, one can focus mainly on the coupling between these $f$-electron states.  If one examines the individual $f$-electron orbitals, one discovers that they have a very complicated arrangement of lobes pointing in many different directions.  From this picture, it is not surprising that any attempt to form the atoms into a periodic array would have difficulty optimizing the overlap of these various $f$ orbitals.  The resulting highly distorted crystal structure appears to be basically some sort of frustrated system of overlaps.  The resulting arrangement of atoms and crystal structure can be calculated and predicted very accurately by full-potential modern first-principles LDA-like band-structure calculations, which automatically finds the lowest total energy (maximum overlap).  Based on this success, it has been suggested that there is very little correlation for $\alpha$-Pu, since standard band-structure calculations have minimal correlation.  Nonetheless, it was noted  quite early on that the volume per atom plots of the actinide series as a function of atomic number showed an upturn for 
$\alpha$-Pu instead of continuing downwards along the parabolic trajectory expected from arguments developed by Friedel~\cite{Friedel:1969,RCAlbers01} based on band-filling effects (this behavior is observed in all the $d$-electron transition-metal elements).  It has often been suggested that this upturn is caused by a small volume expansion in $\alpha$-Pu due to correlation effects. Hence, it is clearly important to examine the amount of correlation in
 $\alpha$-Pu to confirm whether or not this is true.

Because of the large number of atoms in the unit cell, it was previously difficult to apply the most sophisticated electronic-structure techniques to the $\alpha$-phase.  However, a pseudostructure~\cite{JBouchet:2004} was found that approximated quite closely the actual arrangement of atoms in the $\alpha$-structure.  This approximate structure involved an orthorhombic distortion of the diamond structure and had only one nonequivalent type (two sites) of Pu atoms within the unit cell.  This made it possible to perform the only reported LDA+DMFT calculation~\cite{LVPourovskii:2007} of the approximate $\alpha$-phase, which employed the fluctuating exchange $T$-matrix technique for the quantum impurity solver. This calculation showed clear evidence for electronic correlation effects.  Nonetheless, this calculation was not completely satisfactory since it treated each Pu atom in the same way and ignored the uniqueness of the eight different types of different individual atomic sites in the real $\alpha$-structure.  In this respect important physics was lost since the eight different types of atomic sites have significantly different numbers, orientations, and distances of the Pu atoms nearest to them, which causes them to have different competitions between 5$f$ hybridization~\cite{PSoderlind:2004} and onsite Coulomb interactions.  This suggests that each site can have a different amount of relative electronic correlation between each other.

\begin{figure}[b!]
\centering\includegraphics[
width=1.0\linewidth,clip]{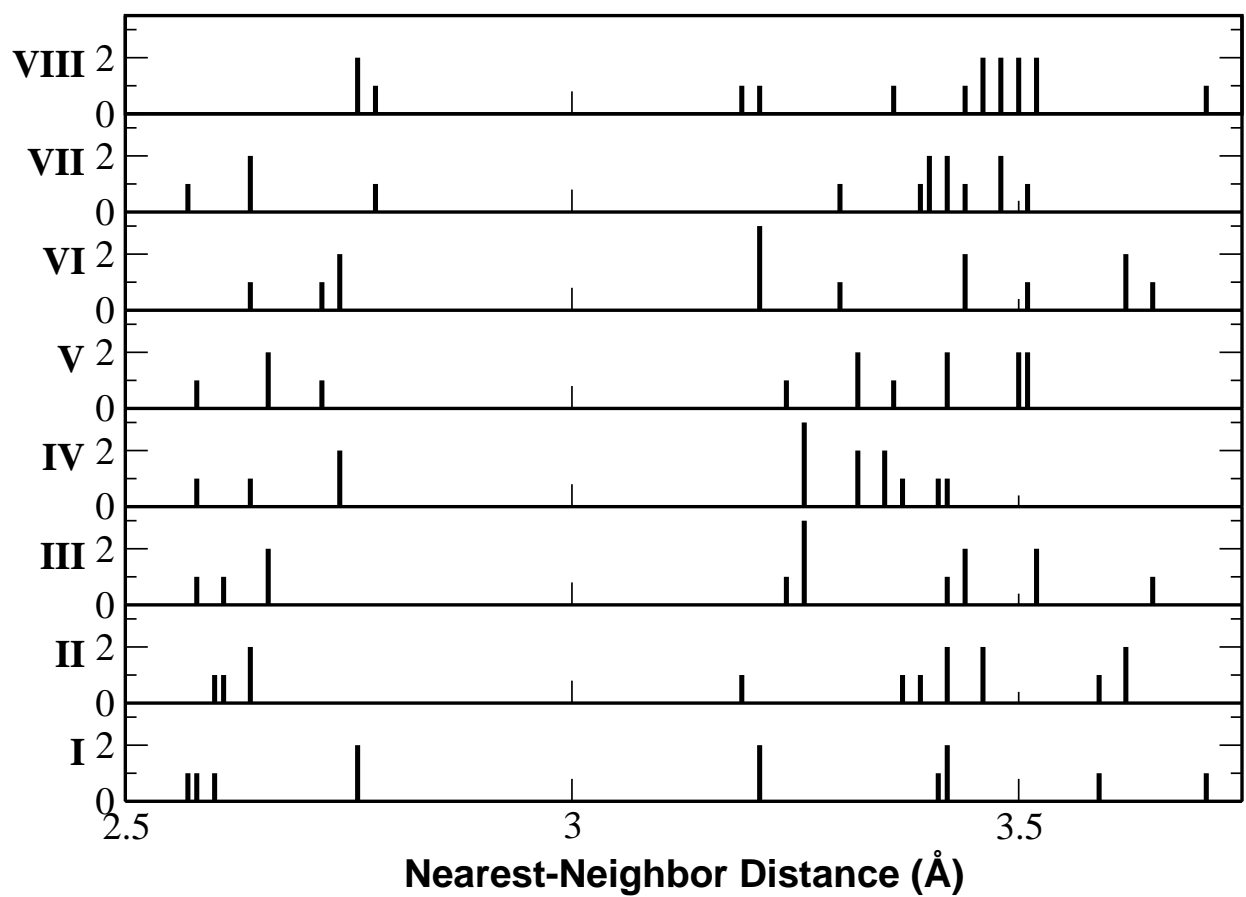}
\caption{{\bf Bond lengths.} Distribution of interatomic distances for eight crystallographically nonequivalent atoms in $\alpha$-Pu.  The height of the lines measures the number of neighbors at the given distance.
}
\label{FIG:nn_pu_sites}
\end{figure}

To explain this in more detail it is useful to plot the distances of the nearest Pu atoms around each site (cf. Fig.~\ref{FIG:nn_pu_sites}).  From these types of plots, it has been found that nearest-neighbor bonds in Pu split into groups of short (between 2.57-- 2.78 \AA) and long (between 3.19 -- 3.71 \AA) nearest-neighbor bond lengths. From Fig.~\ref{FIG:nn_pu_sites} it is quickly evident that the Pu sites can roughly be divided into three categories: Out of eight unique Pu sites, six sites (sites II through VII) have similar types of bonding, with four short nearest-neighbor bonds  and ten long bonds. However, two other sites are qualitatively different.  Site-I has five short bonds with three of them at the shortest of any Pu-Pu atomic distance and seven long bonds,  whereas site-VIII has only three short bonds with all of them at the longest distances in the short-bond range and thirteen long bonds.  Based on this analysis, we would expect that site-I Pu atoms have the strongest hybridization with neighboring atoms (most itinerant), sites II through VII are intermediate, and site VIII is the most localized type of Pu atom.

This suggests, which we confirm with the calculations in this paper, that the $\alpha$-phase of Pu is extremely fascinating from an electronic structure point of view, in that it appears to have the capacity to simultaneously have multiple degrees of electron localization/delocalization of Pu $5f$ electrons within a single material. The Pu atoms on site-I will have itinerant $f$-electron behavior, sites II through VII will be intermediate, and site-VII will be more strongly localized (like the Pu atoms in $\delta$-Pu). Thus an understanding of the electronic structure in $\alpha$-Pu can not only serve as a test of sophisticated theoretical capability but also shed new insight on emergent phenomena in Pu-based compounds~\cite{JXZhu:2012}.

\begin{figure}[t]
\centering
\includegraphics[width=1.0\linewidth,clip]{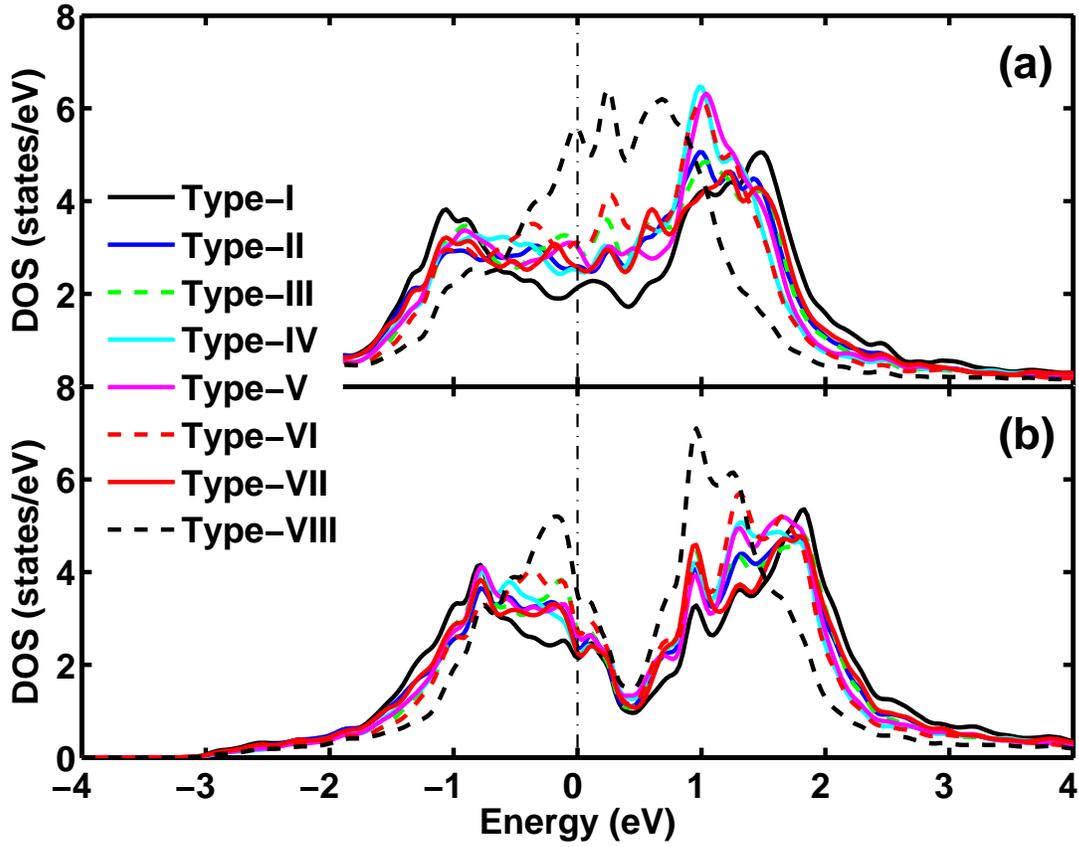}
\caption{{\bf GGA-based Pu $5f$ partial density of states.}   These are shown  for all eight crystallographically nonequivalent atoms in $\alpha$-Pu.  {\bf (a)}  The  spin-orbit coupling is switched off and {\bf (b)} turned on  to show the site-dependence of the DOS. The dashed line in the figure denotes the Fermi energy. 
}
\label{FIG:ldados}
\end{figure}

\noindent
{\bf Electronic Properties.} 
These different local atomic environments have a significant impact on the electronic properties of each individual atom in $\alpha$-Pu.  In Fig.~\ref{FIG:ldados}, we show the calculated GGA band-structure $5f$ partial density of states (DOS) for each of the eight crystallographically nonequivalent sites in $\alpha$-Pu. The results are obtained (a) without and (b) with spin-orbit coupling taken into account. Interestingly,  in the absence of spin-orbit coupling (see Fig.~\ref{FIG:ldados}(a)), the intensity of the DOS at the Fermi energy can be divided into three groups that are exactly analogous to those we found based on nearst-neighbor distances. Out of the eight different atomic types, six of them (sites II through VII) form a group with a similar DOS intensity. In contrast, the type-I site has the smallest intensity while the type-VIII site has the strongest intensity. More remarkably, the DOS for the type-I site exhibits a pseudogap feature. Empirically, a higher Pu $5f$ DOS at the Fermi energy suggests a narrower effective bandwidth of  Pu $5f$ electrons, while a lower intensity at the Fermi energy suggests the opposite.  The appearance of a pseudogap implies the limiting case of strong hybridization between Pu $5f$ and $spd$ conduction electrons.   The strong spin-orbit coupling of Pu (see Fig.~\ref{FIG:ldados}(b)) splits the Pu $5f$ states two manifolds, corresponding to a total angular momentum of $j=5/2$ and $j=7/2$. This spin-orbit coupling effect is most effective for the type-VIII atom but is interwoven with the pseudogap originating from the strong hybridization for the  
type-I atom. This different DOS feature is closely related to the unique atomic environment in $\alpha$-Pu. 

\begin{figure}[t]
\centering
\includegraphics[width=1.0\linewidth,clip]{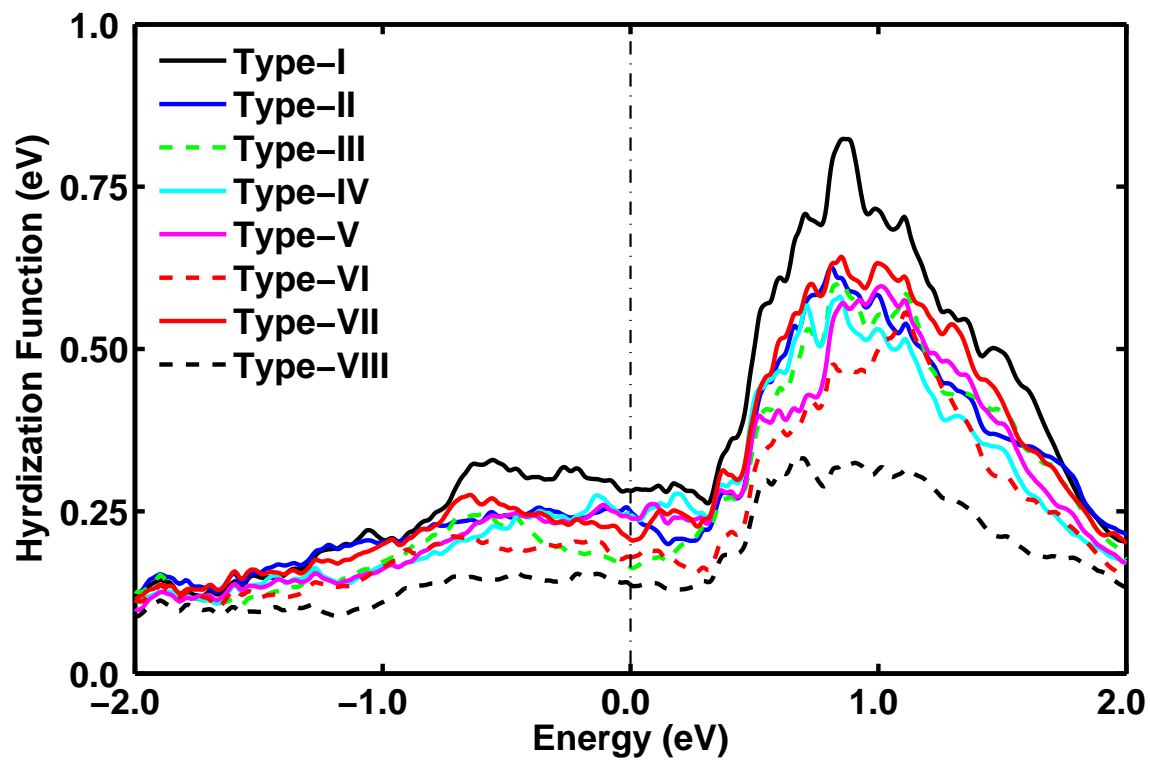}
\caption{{\bf Hybridization strength.} The $j=5/2$-subshell hybridization function for the  Pu $5f$ electrons with the $spd$ electrons at eight crystallographically nonequivalent atoms in $\alpha$-Pu. 
}
\label{FIG:ldahybrid}
\end{figure}

To further clarify the close relation between the electronic structure and the local atomic environment, we evaluate for each site the frequency-dependent hybridization function, $\Gamma(\omega)=-\text{Im}[\Delta(\omega)]/\pi$,  where $\Delta(\omega)$ is the self-energy due to the hybridization of  $5f$ electrons with $spd$ conduction electrons, and is determined by the difference between the inverse Green's functions of the Pu $5f$ electrons in the fully localized limit and that with the Pu $5f$ electrons hybridized. Since the Pu $5f$ electrons mostly occupy the $j=5/2$ subshell, we show in Fig.~\ref{FIG:ldahybrid} the GGA-calculated hybridization function of this subshell.  In a reasonably broad range around the Fermi energy, from about -0.7 eV to about 0.3 eV, it can be seen that the hybridization function is approximately constant in energy. Consistent with the discussion on the nearest-neighbor distances and the GGA-based DOS,  the distribution of the hybridization function strength at the Fermi energy can again be divided into three groups: The type-I and type-VIII sites are the outlier of the distribution with the former having strongest low-energy hybridization strength while the latter having the weakest strength.  Since the hybridization determines exponentially the coherent energy or Kondo energy scale~\cite{ACHewson:1993},  the Pu $5f$ electrons at the type-VIII site are expected to be more localized than those that at the type-I site. Although the LDA hybridization will overestimate the coherent energy scale in crystals as compared to the Kondo impurity problem~\cite{JWAllen:1982}, the trend predicted here is likely to carry over to more sophisticated theories.

\begin{figure}[t]
\centering
\includegraphics[width=0.8\linewidth,clip]{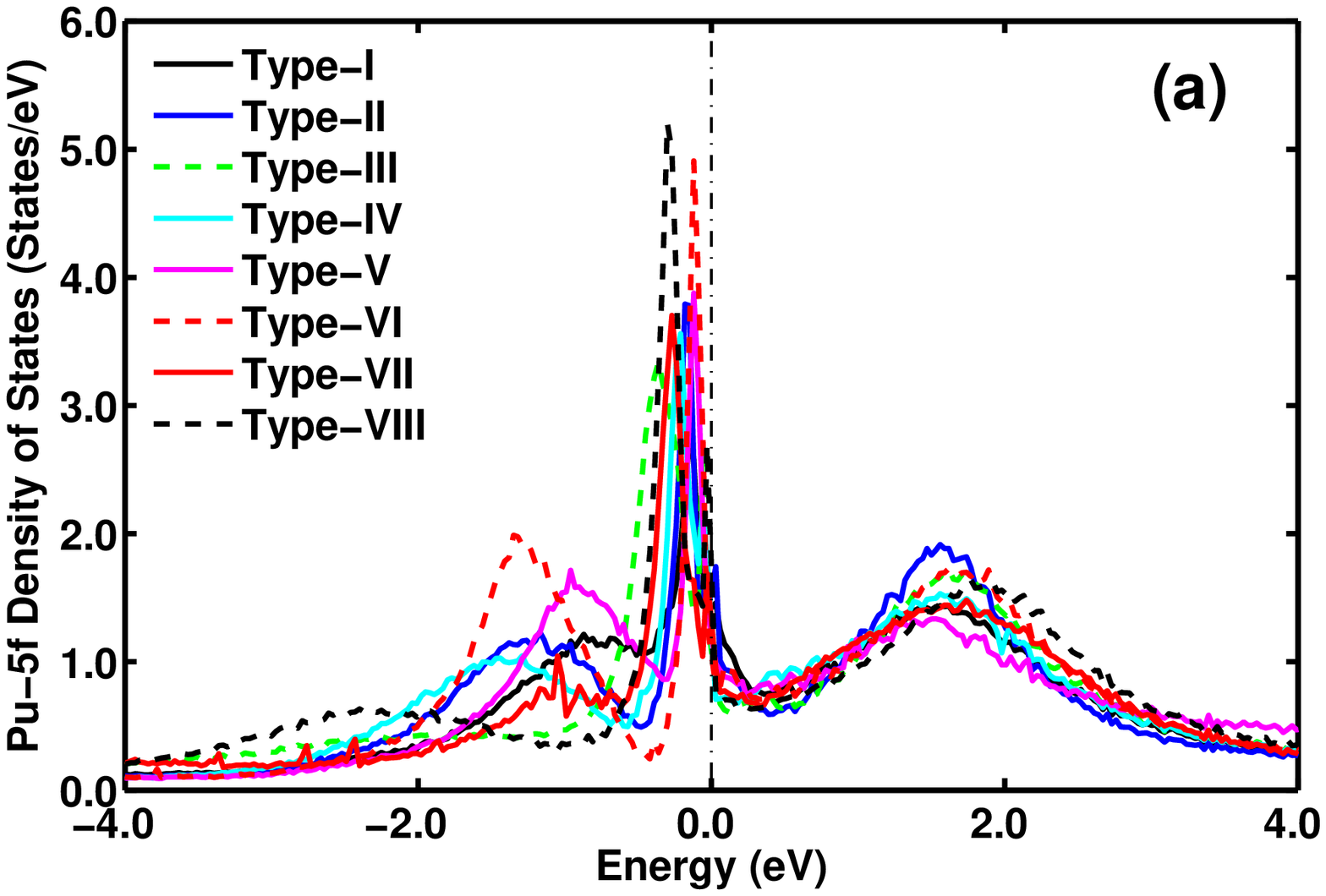}
\includegraphics[width=0.8\linewidth,clip]{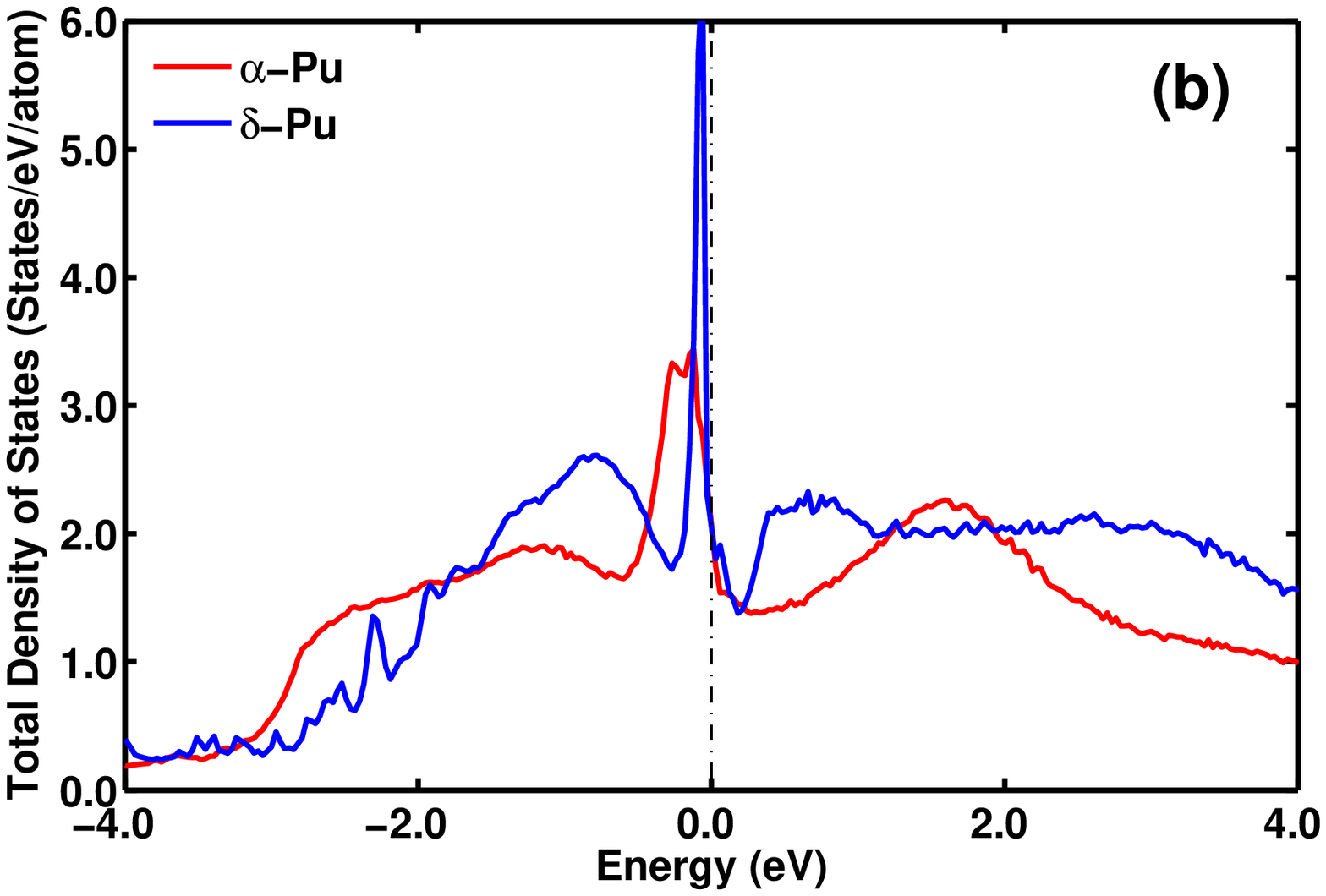}
\caption{{\bf LDA+DMFT-based density of states.}  {\bf (a)} Pu $5f$ partial DOS 
at eight crystallographically nonequivalent sites in $\alpha$-Pu. {\bf (b)} The site-averaged DOS for $\alpha$-Pu in comparison with that for $\delta$-Pu. The calculations are done at a temperature of 232 degrees Kelvin.
}
\label{FIG:dmftdos}
\end{figure}

Figure~\ref{FIG:dmftdos}(a) shows the Pu 5$f$ partial DOS in $\alpha$-Pu. It exhibits a three-peak structure. The two broad peaks below and above the Fermi energy correspond
to the lower and upper Hubbard bands. 
The central peak, which is located very close to the Fermi energy,
is  a coherent quasiparticle state. The appearance of the three-peak structure is 
 a hallmark of quantum many-body effects as revealed in strongly correlated metals like $\delta$-Pu. However, relative to $\delta$-Pu, the $\alpha$-phase electronic structure is much richer. In particular, we observe a very strong site-dependence of the spectral weight transfer on the Pu $5f$ DOS in $\alpha$-Pu. For the six types of atoms with intermediately strong hybridization, the spectral weight is transferred from the lower incoherent Hubbard peak at about $-2$ eV toward the Fermi energy  and forms a new broad peak at about $-1$ eV. This broad peak conspires with the quasiparticle peak at the Fermi energy to exhibit a peak-dip-hump structure, indicative of intermediate electron coupling~\cite{TDas:2012} on these six types of atoms. For the type-I atom,  the broad beak at the lower energy of about $-2$ eV is moved to almost merge with the quasiparticle peak at the Fermi energy and as such the peak-dip-hump structure is smeared out. Furthermore, when we compare the GGA+DMFT-calculated DOS for this type with that calculated with GGA in the presence of spin-orbit coupling, the spectral density at the Fermi energy is similar, except for a strong intensity enhancement at the Fermi energy due to electronic correlation. By contrast,  the Pu $5f$ DOS at the type-VIII site has an appreciable strength of the spectral weight at the lower incoherent band at about $-2$ eV. The spectral feature at this site is very similar to that of very strongly correlated $\delta$-Pu~\cite{CAMarianetti:2008}. 
Finally, we also show in Fig.~\ref{FIG:dmftdos}(b) the averaged DOS per Pu atom in $\alpha$-Pu in comparison with that of $\delta$-Pu. The DOS intensity at the Fermi energy in $\alpha$-Pu is smaller than that for $\delta$-Pu, accompanied by a broader quasiparticle band in the former structure. The pronounced peak-dip-hump structure as observed in $\delta$-Pu is strongly suppressed in $\alpha$-Pu. Experimentally, the result from early PES measurements on the $\alpha$-Pu~\cite{AJArko00} is inconclusive, possibly due to the surface effects. As such, we leave it to the future PES experiments to validate our prediction.

\begin{figure}[t]
\centering
\includegraphics[width=1.0\linewidth,clip]{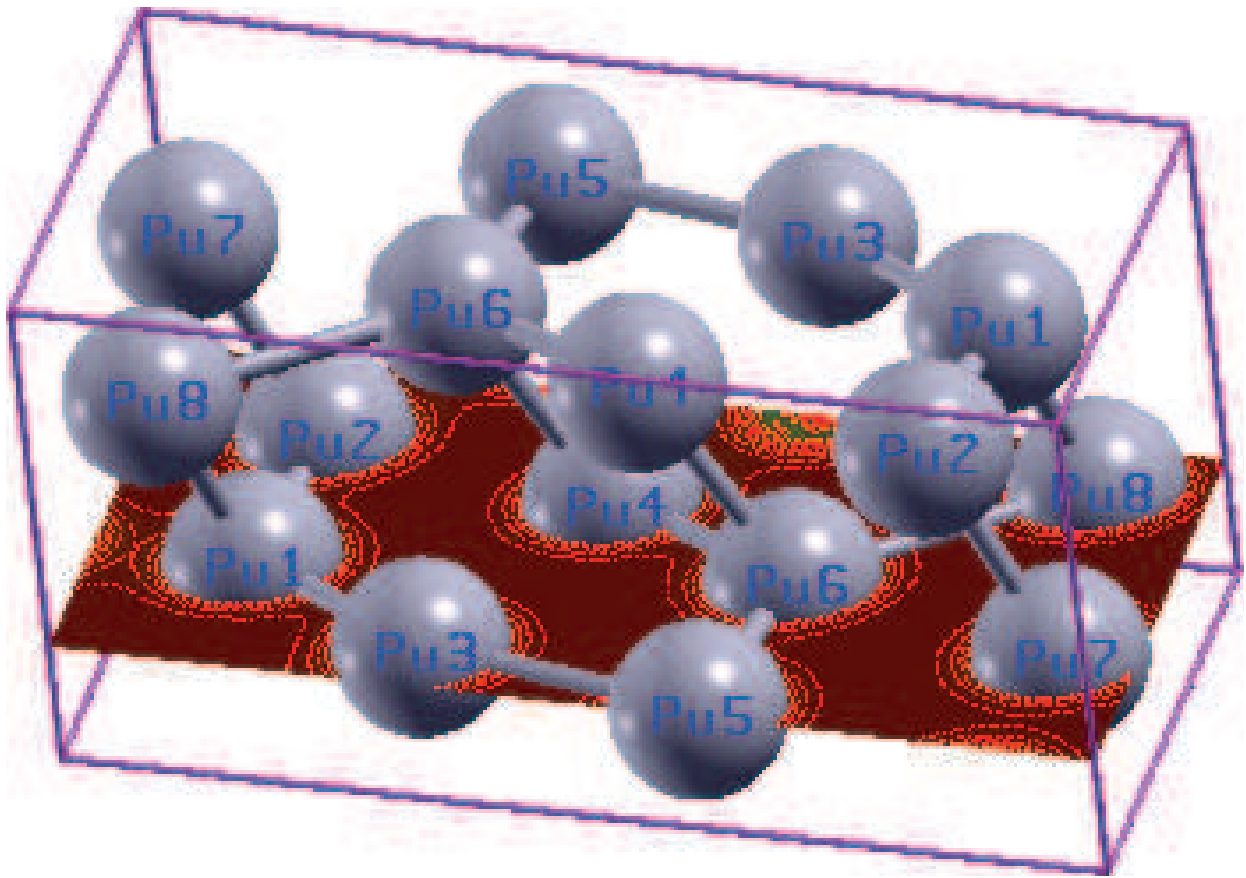}
\caption{{\bf LDA+DMFT-based electron density.}  Electronic charge distribution of $\alpha$-Pu in  (010) plane, which contains 8 nonequivalent atoms using Xcrysden~\cite{AKokalj:2003}. The valence density as represented by contour lines overlaps along the short bonds. 
The calculations are done at a temperature of 232 degrees Kelvin.
}
\label{FIG:electrondensity}
\end{figure}

\noindent
{\bf Discussion}
\\
In addition to the fact that the interatomic distances in $\alpha$-Pu at each site divide into a set of very short bonds and another set of distinctly longer bonds, it is also interesting to note that all the short bonds lie in one hemisphere at each site.  This distribution of the short bonds is closely related to the highly anisotropic Pu 5$f$-electron orbitals (as compared with $s$, $p$, and $d$ orbitals). When these electrons actively participate in the bonding of atoms, which is the case for $\alpha$-Pu, they form a narrow band, implying a significantly large DOS near the Fermi energy. This enhanced Pu-5$f$ DOS causes a re-distribution of electron density in real space, which is shown in Fig.~\ref{FIG:electrondensity}. This phenomenon is reminiscent of the Peierls instability~\cite{PSoderlind:1995}. As a consequence, the enhanced screening of the Coulomb interaction in the electron rich region, together with the unique distribution of the Pu-$5f$ wavefunctions in real space, leads to the agglomeration of short bonds.  We note that these effects are already present at the LDA level, since these calculations predict a minimum theoretical energy at the experimental atomic positions with all the short bonds lying in one hemisphere.  With the additional correlation of the DMFT method, the band narrowing is further amplified. Because of the complex bonding with so many different competing overlaps of the 5$f$ wavefunctions and the anisotropic instabilities caused by the large DOS, it is likely that there is no simple ``hand-waving" argument for this effect.  The fact that theory correctly predicts this hemisphere effect is probably as close as we can get to a simple explanation.

In summary, we have studied the effects of correlation on the electronic structure of $\alpha$-Pu using the LDA+DMFT method. We have uncovered a significant site-selective electronic correlation in this complicated metal. One of the eight Pu sites is fairly itinerant and one is simultaneously very strongly correlated.  The other six are moderately correlated.  The degree of correlation on each site is related to the number and relative distances of the nearest-neighbor atomic positions.  
Sites with a larger number of very short bonds are more itinerant and sites with relatively predominantly longer bonds are more localized.  
Since bond strength is related to correlation strength (correlations weaken bond strength, cf. Refs.~\onlinecite{RCAlbers01,ASvane:2013}), our calculations suggest that some bonds in $\alpha$-Pu will be relatively weaker than other bonds, depending on which sites are connected by the various bonds, leading to even more complex bonding properties than is usual for such a complicated and highly anisotropic crystal structure.  Other physical properties due to the Pu $5f$ orbitals will also similarly exhibit a simultaneous variation in correlation effects.

\noindent
{\bf Methods}
\\
\small
{\bf LDA Calculation Details.}
We first performed electronic structure calculations of $\alpha$-Pu within the framework of density functional theory in the
generalized gradient approximation~\cite{JPPerdew:1996}.  
Our calculations were carried out by using the full-potential linearized augmented plane wave (FP-LAPW) method as implemented in the WIEN2k code~\cite{PBlaha:2001}.
The spin-orbit coupling was included in a second variational
way, for which relativistic $p_{1/2}$ local orbitals were added into the basis set for the description
of the 6$p$ states of plutonium~\cite{Kunes2001}. The muffin-tin radius is 2.31$a_0$ for Pu, where $a_0$ is the Bohr radius. The  $7\times 8 \times 3$ number of $\mathbf{k}$-points was used for the calculations.

\noindent
{\bf DMFT Calculation Details.}
To incorporate the effects of electronic correlations in $\alpha$-Pu, we
then carried out the electronic structure calculations based on
GGA+DMFT method~\cite{GKotliar06}, as implemented in Ref.~\onlinecite{KHaule:2010}. In the GGA+DMFT
method, the strong correlations on Pu atom are treated by the DMFT,
adding self-energy to the DFT Kohn-Sham Hamiltonian. The self-energy
contains all Feynman diagrams local to the Pu atom. No downfolding or
other approximations were used.  Calculations were fully
self-consistent in charge density, chemical potential, the lattice and
impurity GreenÕs functions. The Coulomb interaction $U=4$ eV was used,
consistent with previous work on elemental Pu~\cite{SYSavrasov01,JHShim:2007,JXZhu:2007,CHYee:2010}. The remaining Slater integrals $F^2=6.1$ eV, $F^4=4.1$ eV, and $F^6=3.0$ eV were calculated using Cowan's atomic-structure code~\cite{RDCowan:1981}
 and reduced by 30\% to account for screening. 
 
The impurity model was solved using the strong coupling version of the
continuous time quantum Monte Carlo (CT-QMC) method~\cite{PWerner:2006a,PWerner:2006b,KHaule:2007}, which provides numerically exact solution of the impurity problem. We
recently improved the efficiency of the method by so called lazy trace
evaluations~\cite{CHYee:2012}, which allowed us to perform such large scale
simulations in $\alpha$-Pu. The new algorithm intelligently estimates the
upper and the lower limit for each contribution to the local trace,
and then carries out only a very tiny fraction of the total trace
computation. Typically it is only of the order of 1\% of such
evaluations necessary to determine whether a Monte Carlo move is
accepted or rejected.

The densities of states were computed from analytic continuation of
the self-energy from the imaginary frequency axis to real frequencies
using an auxiliary GreenÕs function and the maximum entropy method~\cite{MJarrell96},
taking care that the zero frequency limit of imaginary and real axis
self-energies agree.

\noindent
{\bf Computational Resources.}
For our calculations we completely dedicated a middle-scale computer
cluster with 24 computer nodes (each with 48 GB of memory and 12 cores)
for more than 2000 non-interrupted wall-clock hours (almost 3 months).  
This involved 10 iterations of charge self-consistency (each containing
10 DMFT iterations and 5 LDA iterations). A large memory was required
since, for example, 4 cores of parallel CT-QMC running on each node used
almost 80\% of the memory of each of our nodes.

\vspace{0.5cm}



\vspace{0.5cm}

\noindent{\bf Acknowledgements~}
We acknowledge useful discussions with T. Durakiewicz, M. J. Graf,  and J. J. Joyce.
This work was performed at Los Alamos National Laboratory under the auspices of the U.S. Department of Energy, and was supported by the LANL LDRD Program (J.X.Z. \& J.M.W.) and the U.S. DOE Office of Basic Energy Sciences under Grant No. DE-FG02-99ER45761 (K.H. \& G.K.).
This work was, in part, supported by the LANL ASC Program.


\vspace{0.5cm}
{\noindent\large\bf Author contributions}\\
J.-X.Z. conceived and designed the study, and carried out the numerical calculations. R.C.A. and K.H. assisted theoretical analysis. All authors contributed to the scientific discussions, and the preparation and discussion of the manuscript. 

\vspace{0.5cm}
{\noindent\large\bf Additional Information}\\
The authors declare that they have no competing financial interests.
Reprints and permission information is available
online at http://www.nature.com/reprints. Correspondence and requests for
materials should be addressed to J.-X.Z. (jxzhu@lanl.gov).


\newpage
\noindent {\bf FIGURE LEGENDS}

\end{document}